\documentclass{article}

 \usepackage[preprint, nonatbib]{neurips_data_2021}

\usepackage[utf8]{inputenc} 
\usepackage[T1]{fontenc}    
\usepackage{hyperref}       
\usepackage{url}            
\usepackage{booktabs}       
\usepackage{amsfonts}       
\usepackage{nicefrac}       
\usepackage{microtype}      
\usepackage{graphicx}
\usepackage{color}
\usepackage{makecell}
\usepackage[table,xcdraw]{xcolor}
\usepackage[tableposition=top]{caption}
\usepackage{tabularx}
\usepackage{subcaption}
\usepackage{adjustbox}

\title{JS Fake Chorales: a Synthetic Dataset of Polyphonic Music with Human Annotation}

\author{%
  Omar A.~Peracha \\
  Humtap, Inc. \\
  London, U.K. \\
  \texttt{omar@humtap.com} \\
}

\begin{document}

\maketitle

\begin{abstract}
High-quality datasets for learning-based modelling of polyphonic symbolic music remain less readily-accessible at scale than in other domains, such as language modelling or image classification. Deep learning algorithms show great potential for enabling the widespread use of interactive music generation technology in consumer applications, but the lack of large-scale datasets remains a bottleneck for the development of algorithms that can consistently generate high-quality outputs. We propose that models with narrow expertise can serve as a source of high-quality scalable synthetic data, and open-source the JS Fake Chorales, a dataset of 500 pieces generated by a new learning-based algorithm, provided in MIDI form. 

We take consecutive outputs from the algorithm and avoid cherry-picking in order to validate the potential to further scale this dataset on-demand. We conduct an online experiment for human evaluation, designed to be as fair to the listener as possible, and find that respondents were on average only 7\% better than random guessing at distinguishing JS Fake Chorales from real chorales composed by JS Bach. Furthermore, we make anonymised data collected from experiments available along with the MIDI samples. Finally, we conduct ablation studies to demonstrate the effectiveness of using the synthetic pieces for research in polyphonic music modelling, and find that we can improve on state-of-the-art validation set loss for the canonical JSB Chorales dataset, using a known algorithm, by simply augmenting the training set with the JS Fake Chorales.
\end{abstract}
\section{Introduction}\label{sec:introduction}

\begin{figure}
 \centerline{
 \includegraphics[width=0.35\columnwidth]{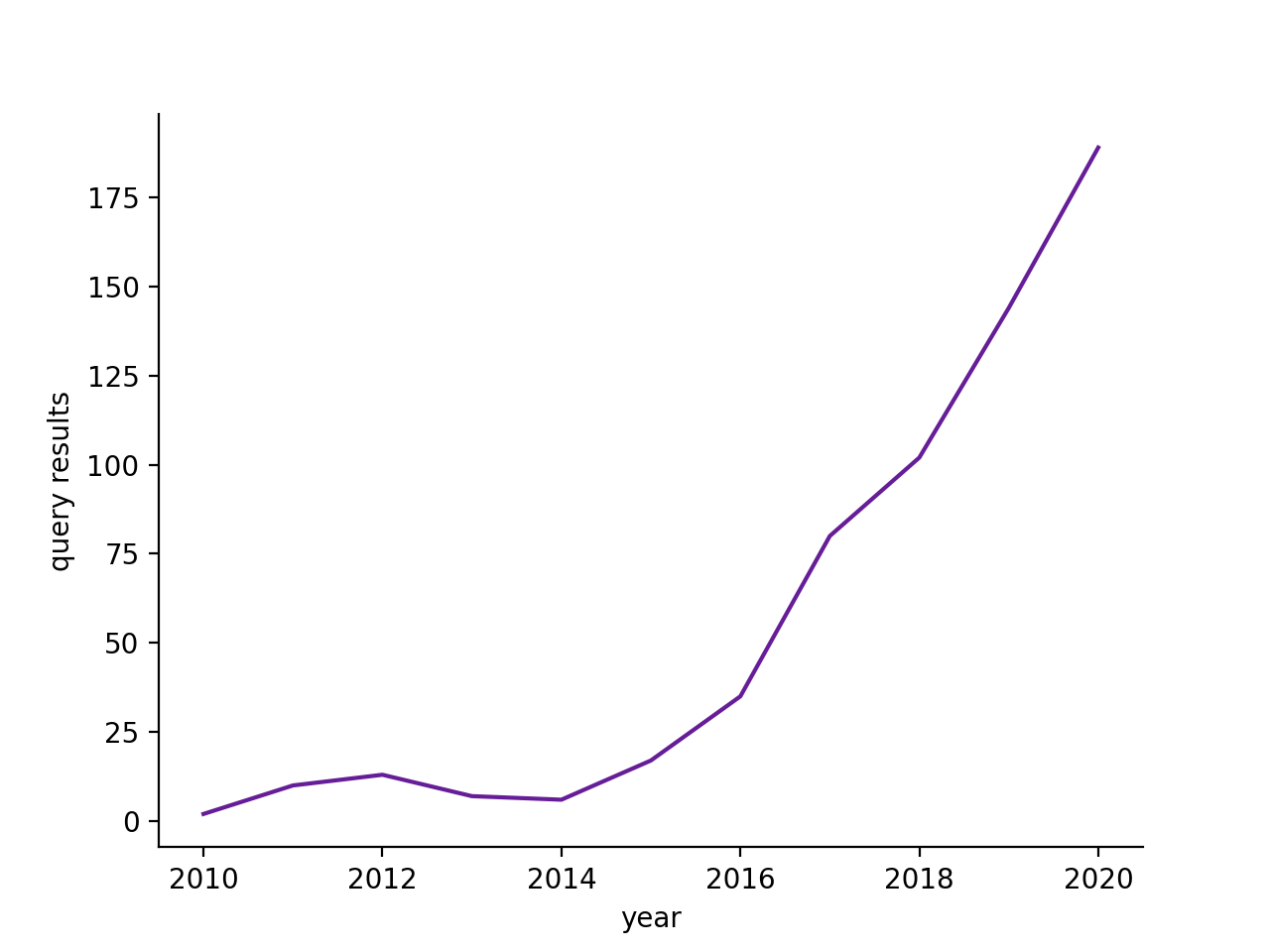}}
 \caption{Number of results returned by \cite{arxiv} when filtering for papers in the Computer Science, Mathematics or Statistics categories with both "Music" and "Learning" in their abstracts for the years 2010-2020 inclusive. For years before 2015, we further filter out any irrelevant results manually.}
 \label{fig:musicai}
\end{figure}

Representation learning for application to music has become an increasingly active field of research in recent years, as Figure \ref{fig:musicai} demonstrates. Despite growth in the domain, certain bottlenecks have persisted which hinder the rate of breakthroughs, particularly regarding problems related to symbolic music. The most obvious of these concerns the volume of high-quality datasets available for researchers to work with cheaply and with low effort; datasets comprising millions of samples \cite{imagenet} or even tens of billions of canonical input/output pairs \cite{books} have been available for many years in the fields of computer vision and natural language processing. Some of these are even licensed with sufficient permissiveness as to allow commercial use of any technology whose development relied upon said data \cite{openimages}, a factor which plays an important role in determining the ultimate viability of deploying these algorithms into the real world where they may generate value for businesses and customers alike, naturally promoting further investment. Unsurprisingly, the aforementioned two domains continue to see incredible progress in research and increasingly wide adoption of the resulting technology in consumer-facing products.

A second bottleneck more pertinent to generative music modelling is the lack of automated methods of assessing outputs which strongly correspond to good qualitative performance. This difficulty in measuring success naturally leads to slower progress. Few alternatives have been proposed to human evaluation, which is typically slow or costly, for assessing the quality of algorithmically-generated music. The process of acquiring human feedback itself tends to vary in implementation details between different published works \cite{deepbach, bachbot, coconet}, which further convolutes the process of reliably comparing results in the literature.

We propose a dataset of synthetic polyphonic music, generated by our neural network-based \textit{KS\_Chorus} algorithm, in symbolic form, and aim to validate both its usefulness for aiding in symbolic music modelling tasks and its ability to scale on demand. We achieve the latter by building the dataset from 500 consecutive generated outputs of \textit{KS\_Chorus}, with zero cherry-picking of samples, and performing an experiment designed to solicit human evaluation in a context entirely free from curation, which may otherwise serve to skew results in the algorithm's favour. 500 samples is small enough that conducting such an experiment remains fairly manageable, but is large enough relative to the set of 382 chorales by Johann Sebastian Bach \cite{boulanger}, whose style the JS Fake Chorales seek to imitate, that the results may be deemed significant. After 6,810 responses collected from an estimated 446 unique respondents, our experiment shows that human listeners are only 7.3\% better than a coin flip at telling apart our synthesised samples from chorales composed by Bach himself. We facilitate scalability by making a web app freely available for researchers to generate new samples from \textit{KS\_Chorus},\footnote{Web app to generate new samples is available at \url{https://omarperacha.github.io/make-js-fake/}} enabling them to grow the dataset, which is further described in Section \ref{subsec:webapp}.

We then conduct experiments using the JS Fake Chorales as training data or auxiliary data on a common benchmark task, namely modelling the canonical JS Bach Chorales dataset. Using the current state-of-the-art model, \textit{TonicNet} \cite{tonicnet}, without any tuning of parameters, we show that training purely on JS Fake Chorales achieves performance close to state-of-the-art on the JS Bach Chorales in terms of validation set loss, even without training on any pieces from the actual JS Bach training set. By combining the JS Bach Chorales training set with the JS Fake Chorales, using the latter as a form of dataset augmentation, we can improve on  state-of-the-art results. We make all code to run our ablation studies publicly available, for reproducibility.\footnote{Code to reproduce results from experiments is available at \url{https://github.com/omarperacha/TonicNet}}

In evidencing the scalability of the JS Fake Chorales and their usefulness in common music modelling tasks, we hope to highlight the possibility for synthetic data to help alleviate the current bottleneck of low data volume in this domain. Finally, to help address the bottleneck of weak methods for measuring qualitative performance, we make anonymised metadata for each of the 6,810 human responses available as part of the dataset,\footnote{Dataset and usage information are available at \url{https://github.com/omarperacha/js-fakes}} which includes information about the evaluator's level of music education and which specific pieces were guessed correctly or incorrectly by which evaluator, among other data further detailed in Section \ref{sec:dataset}.  Future work may see attempts at modelling these data as a means to automatically infer qualities which make music more readily-perceived as human-composed.

Ours is the first dataset in this domain that provides such data for such a large number of samples without cherry-picking, to our knowledge. Similarly, Section \ref{subsec:humaneval} shows that \textit{KS\_Chorus} is the first deep learning algorithm proven to consistently approach, though not yet achieve, a production-level consistency of high-quality unconditioned outputs in Bach's, or indeed any  clearly-defined polyphonic, style. The value of this is that the generated samples can already be very useful as auxiliary training data in downstream tasks, as shown in Section \ref{subsec:algoeval}.

In a data-starved domain, models with narrower focus are able to squeeze performance on specific genres from the lack of a requirement to generalise to other styles, for example by making style-specific adjustments in the representation. As a concrete example, \textit{KS\_Chorus} does not need to handle individual voices capable of producing more than one note simultaneously, allowing for a more efficient representation of Bach's chorales than cross-genre polyphonic music models, such as \textit{MuseNet} \cite{musenet}. We propose that narrow-focussed models, by generating high quality synthetic data in a single or small number of styles, may play a role in significantly increasing the availability of training data, thus empowering future generative algorithms to generalise better across styles while improving on both quality and consistency.

\section{Related Work}\label{sec:relatedwork}
Synthetic datasets in the music domain have so far seen wider use in audio contexts rather than symbolic music. For example, both \cite{pyin} and \cite{crepe} trained algorithms on synthetic data to learn how to accurately detect the fundamental frequency of monophonic musical material. While algorithms which process audio have seen more benefit from synthetic data up until now, examples do exist of such data playing a role in symbolic music contexts; in \cite{gankyoku}, extra training data is synthesised by performing a conditional nonlinear transform on the original training corpus, in a naive 1-dimensional token embedding space. This data is used to augment the training set and is fed to the generator and discriminator in a generative adversarial learning context, with an input label corresponding to the intensity of the transform performed on the original data sample so as to allow both networks to learn features expressed in the extra data which are also present in the original data. It is worth noting that the intent in this case was not to create perfect stylistic imitations of the original dataset, but to introduce novelties in perceivably-systematic ways. Nonetheless, \cite{gankyoku} serves as a precursor for using synthetic data to improve the quality of generated symbolic music.

With 10,854 complete pieces of solo piano music in MIDI form, \cite{giantmidi} is among the larger symbolic music datasets available. It could be considered a synthetic dataset in that the authors used an automated method of transcription \cite{piano} to convert over 1000 hours of audio recordings into a symbolic representation. The authors report validation onset and offset F1 score of 0.825, and a score of 0.967 for onsets alone. These are impressive results in the current landscape, yet clearly this may translate to many transcription errors across a dataset comprising millions of note events. The severity of these inconsistencies with respect to the stylistic integrity of the original human-composed piece can be hard to measure, and further work is required to determine the areas where this dataset will ultimately enable progress in the field.

Both \cite{deepbach} and \cite{bachbot} propose algorithms intended to generate stylistic imitations of Bach's chorales, and use human evaluation to support claims of the effectiveness of their solutions. The experiment conducted by \cite{deepbach} first takes 50 pieces from the validation set of the Bach chorales and uses their \textit{DeepBach} algorithm to reharmonise them. They repeat the process with other baseline algorithms. They then take 12-second extracts of each and play these samples one-by-one, asking listeners to determine if they think each is by Bach or is a generated piece. In contrast, we generate 500 pieces entirely from scratch with no conditioning and always play the entire piece to listeners, which can be over one minute in length. We also make a ground truth chorale by Bach available to the listener to play back at any time alongside the sample currently being evaluated.

The experiment conducted by \cite{bachbot} is similar to ours in that it allows to listen to two samples simultaneously. Unlike ours, however, the listener must simply pick which of the two samples they feel is more Bach-like. In our case, the evaluator knows that one sample is definitely composed by Bach, and must decide if the second sample is also by Bach or if it is generated by \textit{KS\_Chorus}. A second significant difference is that \cite{bachbot} only use a set of 12 compositions fully-generated by their \textit{BachBot} algorithm in their experiments. This is too small a sample to substantiate the model's general performance. In our case, we generate 500 samples in a row and include all of these in the human evaluation experiment without cherry-picking. Neither \cite{deepbach} nor \cite{bachbot} make a substantial dataset of generated compositions available, while we make these initial 500 available as the JS Fake Chorales dataset, and offer evidence that the dataset can be further scaled while maintaining the same quality consistently as a result of avoiding curation during human evaluation.

Perhaps the most relevant work to ours is the Bach Doodle Dataset \cite{bachdoodle}, obtained by allowing users to interact with the \textit{CoCoNet} algorithm \cite{coconet}; the user could enter a melody via a web interface, and the algorithm would harmonise this melody in the style of Bach. Comprising 21.6 million samples, the size far exceeds any that is commonly-used in symbolic music, to our knowledge. The web interface also collects a user rating on a three-point scale to reflect their satisfaction with the output. This rating can not be said to reflect stylistic consistency with any intended target, however, but simply provides a general measure of the user's experience. This rating is provided as part of the dataset, along with more metadata about the user's session. Because the melodies are user-entered, only constrained to fit within a selected key and within a 2-bar duration, the quality of the final samples in the Bach Doodle Dataset will be inherently unpredictable. Furthermore, all samples are fixed to 2 bars in duration, and melodies are limited to quaver resolution. Our samples consist only of pieces in their entirety, with a median bar duration of 11.375 and maximum of 35.5, and rhythmic resolution can fully capture that seen in the original Bach chorales. 

Several style-agnostic polyphonic models have been proposed in recent years. Three examples notable for being particularly high-performing are \textit{MuseNet} \cite{musenet}, \textit{MMM} \cite{mmm} and \textit{MusicBERT} \cite{musicbert}. While \textit{MMM} and \textit{MusicBERT} are highly promising, \textit{MuseNet} is the only one of these to be evaluated on full song generation. This evaluation is not performed quantitatively, and provided samples are cherry-picked, though an interface for novel song generation is available. It is likely that all three models have been trained on the Bach Chorales, though not known for certain as not all the datasets are public. In any case, none of these models are able to demonstrate the capability to generate a dataset with quality matching the JS Fake Chorales. 
\section{Dataset}\label{sec:dataset}

\subsection{Synthesising 4-part Chorales}\label{subsec:synthesising}

To create the JS Fake Chorales dataset, we first develop a learning-based sequence modelling algorithm, \textit{KS\_Chorus}, and train it on the Bach chorales as made available by the music21 toolkit \cite{music21}. Bach's chorales are a good choice as the basis for a synthetic dataset mainly because they are widely-cited in the literature regarding algorithmic modelling of polyphonic music \cite{deepbach, bachbot, coconet, boulanger, tonicnet}, and so downstream quantitative analysis against good benchmarks is readily possible. 

\textit{KS\_Chorus} is a generative algorithm for polyphonic music of any number of instruments, where each instrument is itself monophonic. It consists of an ensemble of deep RNN-based networks with some domain-specific architectural additions, totalling roughly 130 million parameters, and a sampling routine idiosyncratic both to this architecture and the data representation utilised. We do not detail the specifics of the architecture, input representation or training process here, and instead leave this to a separate work for the future.

Once trained, we then sample 500 compositions from the model consecutively and fully unconditionally, i.e using absolutely no priming or similar input to specify parameters such as the length of the compositions. The only exception to this relates to metre; while a small handful of pieces exist in the original training corpus which are not in 4/4 time, we found during development that \textit{KS\_Chorus} falls somewhat below par when generating compositions in other time signatures. We therefore manually restrict all generated samples to be in 4/4, because the likelihood of being correctly identified as algorithmically-composed seemed to be consistently higher in other time signatures.

Each generated composition is compared to each of the 344 Bach chorales in the training set seen by the model during development to ensure originality; in particular, the edit distance to every single sample in the training corpus must be greater than 50\%, measured separately across both the entire piece and across the opening few bars. An edit distance below 50\% in either case would invalidate the sample and trigger the algorithm to run once more. Furthermore, all samples in the original dataset were transposed as far as possible in either direction while maintaining a singable range for each voice, and the edit distance was similarly validated to these transpositions. A single composition takes roughly 10 minutes on average to generate when running on CPU. The algorithm was implemented and trained using the PyTorch library \cite{pytorch}, and sampling was also performed in Python.

While we did not measure the edit distance of generated pieces to all other generated pieces at sample time, we perform post-hoc analysis of this metric. Since we do not detail the representation used by \textit{KS\_Chorus} in this work, for the sake of reproducibility we perform these comparisons using a standard representation seen in the literature \cite{tonicnet, musictransformer}; the piece is split into 16th-note time-steps and represented as a matrix $x \in \mathbb{Z}^{4\times T}$, where $T$ is the number of time-steps in that sample, the four rows represent the four different voice parts and the value of each element is the pitch observed in the respective voice at the respective time-step. This matrix is typically converted into a linear sequence before being input to a learning algorithm, by end-concatenating the columns. We also make the JS Fake Chorales available in this representation.

Using the above representation, we compare the intra-set edit distance (i.e. comparing each sample with every sample from the same dataset) for both the Bach chorales and the JS Fake Chorales, and we repeat the comparison of each JS Fake chorale with each Bach chorale. Table \ref{tab:distance} shows the results of taking the smallest edit distance from each sample in dataset 1 to each sample in dataset 2 for the three different dataset configurations, and computing the mean of these smallest distances. We also provide the minimum in each case. We can see that there is in fact more intra-set similarity among the Bach chorales than among the JS Fakes. Of course, the edit distance between the two was enforced at generation time, hence the distribution of inter-set least edit distances trending much higher than that of either intra-set least edit distances. We can deduce from this post-hoc analysis that 50\% was perhaps an unnecessarily strict threshold for edit distance during the process of sampling from \textit{KS\_Chorus}.

\begin{table}
\centering
\caption{Results of taking the smallest edit distance from each sample in dataset 1 to each sample in dataset 2 and computing both the mean and the minimum of these smallest distances.\newline}
\label{tab:distance}
\begin{tabular}{rrrr}
\multicolumn{1}{l}{\textbf{Dataset 1}} & \multicolumn{1}{l}{\textbf{Dataset 2}} & \multicolumn{1}{l}{\textbf{Minimum}} & \multicolumn{1}{l}{\textbf{Mean}}  \\ 
\hline
JSF                                    & JSB                                    & 55.078\%                             & 72.818\%                           \\
JSF                                    & JSF                                    & 25.625\%                             & 66.657\%                           \\
JSB                                    & JSB                                    & 12.755\%                             & 57.792\%                          
\end{tabular}
\end{table}

\subsection{Human Annotation}\label{subsec:annotation}

We conduct a human evaluation experiment hosted on the internet to investigate how these generated compositions are perceived in comparison to Bach's chorales. We first present the experiment design in this section and describe the data obtained and provided as part of the JS Fake Chorales dataset, while the actual results of this experiment will be discussed in Section \ref{sec:results}. We use $\mathcal{G}$ to notate the set of 500 pieces which comprise the JS Fake Chorales dataset, and $\mathcal{C}$ to refer to the training corpus of 344 chorales by J.S. Bach.

\subsubsection{Listening Experiment}

Participants are first prompted to self-report their highest level of musical education, selecting from a set of six coarse pre-defined options: 0, no musical education; 1, some musical education, but no formal qualifications; 2, high School level qualifications directly related to music; 3, undergraduate degree directly related to music; 4, postgraduate degree directly related to music; and 5, postgraduate degree specialising in the music of J.S. Bach.
  
Once they have confirmed their choice, they are taken to a page with instructions describing how to complete the experiment. On this page, the participants are provided with a reference chorale by Bach to listen to, chosen randomly per session from the 344 pieces $[C_{1},...,C_{344}] \in \mathcal{C}$. Participants are informed that the given chorale is a genuine example of Bach's music, and they must play it through to the end in order to advance further. Upon doing so, a second piece will appear, which is selected at random from the combined set of all pieces $\mathcal{G} \cup \mathcal{C}$ (excluding whichever Bach chorale was used as the reference), with equal probability of any sample in this set being selected.

Participants are not informed whether this second piece is composed by Bach or \textit{KS\_Chorus}, but are tasked to determine this. Once more they can not proceed without listening to the given sample in its entirety, at which point a menu will appear through which they can provide their response and submit. All pieces are rendered using the same piano synthesiser, and can be played as many times as the participant wants before submitting a response, including the reference track. \textit{KS\_Chorus} does not currently include tempo data as part of the representation it learns to model, therefore all piece are played back at a fixed tempo of 120bpm.

Upon submitting a response, participants are taken to a new page where they are immediately shown the correct answer, and given the option to try another round. They are also shown a running score which increments by 1 for each correct response in the session, which carries over if they choose to try a new sample. Should they choose to do so, they will be taken back to the previous page where the same reference piece will be available for them to play back as desired, and a new sample will be loaded for the participant to identify as belonging to $\mathcal{G}$ or $\mathcal{C}$, ensuring this new sample has not previously been seen in the current session. In this case, participants are not forced to replay the reference track, but must once more listen through the new sample entirely before proceeding.

Responses were solicited both organically via social media, and through Amazon MTurk \cite{mturk}. In the latter case, participants were required to obtain a score of 10 in order to fulfil the conditions for payment, and were given a 30 minute limit to do so. This was done to motivate participants to try their utmost to provide correct responses; mandating that the sample be played through entirely for each round vastly increases the time cost for every wrong answer, meaning that random guessing is not an effective strategy for MTurk respondents to earn the available reward. The mean time taken by MTurk workers to complete the task was 19m45s, and we estimate these individuals make up roughly 66\% of all unique respondents. Overall we received 6,810 responses from 446 unique IP address hashes. Responses were collected throughout the month of February 2021, and the website remains live for anyone to interact with,\footnote{Listening test available at \url{https://omarperacha.github.io/listening-test}} while data is no longer being collected from sessions.

\subsubsection{Metadata}

Besides a simple binary value denoting the accuracy of a participant's response, we collect several other data points designed to offer insight on how complex a given sample was to identify by the participant. Examples include how many times the sample was played before submitting a response and how long the participant took to submit once they had heard the sample for the first time. We aggregate these and make them available as part of the dataset in the hope that this information might provide value to researchers.

For each generated chorale $G_{i} \in \mathcal{G}$, we make four additional pieces of metadata available: (1) Total responses concerning sample $G_{i}$; (2) responses which correctly identified $G_{i}$ as composed by an algorithm; (3) mean number of times the sample was played before a response was submitted; and (4) mean time in seconds between hearing the sample and submitting a response. We provide these data separately for each skill level, to provide further granularity on how the samples were perceived by different demographics. This allows such custom analysis as examining which pieces were likely to be misidentified by people with university degrees in music, and which pieces did people with limited musical education find simplest to correctly identify. We also provide the data aggregated over all skill levels. Further usage explanation is provided in the documentation.

\subsection{Web Application for Generating New JS Fake Chorales}\label{subsec:webapp}

We make a web application freely available for anyone to sample new chorales with \textit{KS\_Chorus}. The application consists of a single page, including text and a button to trigger sample generation. Once the sampling process is complete, the button is replaced with a MIDI player where the user can preview the newly-generated sample, rendered with a piano synthesiser. The MIDI file will also be automatically downloaded to the user's local hard drive from their browser at that time. The main JS Fake Chorales repository will be periodically updated to include all samples generated from the web app, and users are informed of this. No user data of any kind is captured from this interface, besides the generated sample itself. While we commit to the persistency of the JS Fake Chorales dataset, and all samples generated from this web app in future included therein, we may in due course choose to discontinue the availability of this app as we see fit, for example if the cost of upkeep is deemed no longer sustainable.

\section{Experimental Results}\label{sec:results}

\subsection{Human Evaluation}\label{subsec:humaneval}

\subsubsection{Sample Category Identification Task}

\begin{table}
\centering
\caption{Number of samples by Bach and by \textit{KS\_Chorus} seen over all participant sessions during human evaluation (columns 2 \& 5), the number of times these were correctly identified (columns 3 \& 6) and the resulting rate of correct responses (columns 4 \& 7), shown across all skill levels.\newline}
\label{tab:humaneval}
\begin{tabular}{rrrrrrrr}
\multicolumn{1}{l}{\textbf{Skill}} & \multicolumn{3}{c}{\textbf{Bach}}                                                     & \multicolumn{3}{c}{\textbf{A.I.}}                                                     & \multicolumn{1}{l}{\textbf{Total Samples}}  \\
\multicolumn{1}{l}{}               & \multicolumn{1}{l}{Samples} & \multicolumn{1}{l}{Correct} & \multicolumn{1}{l}{Ratio} & \multicolumn{1}{l}{Samples} & \multicolumn{1}{l}{Correct} & \multicolumn{1}{l}{Ratio} & \multicolumn{1}{l}{}                        \\ 
\hline\hline
0                                  & 705                         & 339                         & 48.1\%                    & 1081                        & 633                         & 58.6\%                    & 1786                                        \\
1                                  & 943                         & 445                         & 47.2\%                    & 1351                        & 748                         & 55.4\%                    & 2294                                        \\
2                                  & 515                         & 270                         & 52.4\%                    & 732                         & 409                         & 55.9\%                    & 1247                                        \\
3                                  & 264                         & 98                          & 37.1\%                    & 366                         & 225                         & 61.5\%                    & 630                                         \\
4                                  & 215                         & 97                          & 45.1\%                    & 348                         & 223                         & 64.1\%                    & 563                                         \\
5                                  & 120                         & 65                          & 54.2\%                    & 170                         & 83                          & 48.8\%                    & 290                                         \\ 
\hline
\multicolumn{1}{l}{ALL}            & 2762                        & 1314                        & 47.6\%                    & 4048                        & 2321                        & 57.3\%                    & 6810                                       
\end{tabular}
\end{table}

We compare the likelihood of participants determining a sample to be human-composed, and present results across skill levels for samples by both Bach and \textit{KS\_Chorus}. Given an ideal model whose distribution of generated compositions $\widetilde{G}$ exactly equals the distribution of training samples $\widetilde{C}$, we would expect the likelihood of any $G_{i} \in \mathcal{G}$ and $C_{i} \in \mathcal{C}$ being deemed as human-composed to also be equal. Given a fair experiment which does not incur bias, for example by disproportionately highlighting certain samples of Bach's pieces or suppressing specific qualities present in $\mathcal{C}$ from being observed by the evaluators, this expected likelihood should be 50\%.

Table \ref{tab:humaneval} shows the proportion of all responses which correctly identified samples from $\mathcal{G}$ and $\mathcal{C}$, broken down by skill level. We see that the total rate of correct responses for $\mathcal{C}$ is 47.6\%, while for $\mathcal{G}$ it is 57.3\%. These values, though unequal, are close to each other, and in particular they suggest evaluators did not differ greatly from random chance regarding the likelihood of a correct guess for either corpus. This is despite receiving instant feedback after each round and a reference $C_{i} \in \mathcal{C}$ being made accessible at all times while prompting for a response, which should both aid in increasing the participants' success rate.

We perform Fisher's exact test on a contingency table consisting of rows $\mathcal{C}$, $\mathcal{G}$ and columns \textit{voted Bach}, \textit{voted A.I.} across all skill levels to interpret whether samples from $\mathcal{G}$ did indeed have a similar likelihood to those from $\mathcal{C}$ of being considered human-composed. We test each skill group separately, the combination of all skill groups together, and two subgroups dividing the respondents into those who self-reported holding a formal qualification in music, and those who did not. We also compute the F1 score for each group as a measure of overall performance on the classification task. We report these results in Table \ref{tab:stats}, and see that at the 99.9\% critical value we can accept the null hypothesis that $\mathcal{C}$ and $\mathcal{G}$ are equally likely to be voted as being written by Bach, for every individual skill group and for the two larger subgroups. However, the value of $p$ varies greatly across these groups, and for the overall sample comprising all responses $p=6.432e^{-5}$, so we must reject the null hypothesis. Therefore while at a high level the JS Fakes are certainly difficult to distinguish from Bach Chorales, performing further tests makes it clear there is room for improvement in the quality of the generated samples. Closing this gap is precisely an example of an application which this dataset might serve in future.

We can also see from Table \ref{tab:stats} that there is no apparent trend in F1 score across respondent skill levels, which seems surprising. One obvious candidate for this is the inherent noise and unreliability introduced by enabling respondents to self-report their skill. A stronger method for a future experiment would involve assessing user skill level via some short test. However, the fact that $p>0.001$ consistently across all skill groups suggests participants within each group did in fact respond similarly to their peers. Another criticism we can make is that while likelihood of considering a piece from either sample as human-composed was similar, this observed value was not 50\%, but 55.7\% as measured by weighted average recall. While we have theories for why this may be the case, we must conclude that there are likely some aspects of the experimental design which are not free from introducing small bias. We leave the experiment fully accessible online for transparency and leave it for future work to analyse and improve on the fairness of designing such an experiment.

\begin{table}
\centering
\caption{Results of performing Fisher's exact test and computing F1 score for each skill group, all skill groups combined, and two subgroups dividing the respondents into those who do and do not hold a formal qualification in music. \newline}
\label{tab:stats}
\begin{tabular}{l|rrrrrrrrr}
\textbf{SKILL}      & \multicolumn{1}{l}{0} & \multicolumn{1}{l}{1} & \multicolumn{1}{l}{2} & \multicolumn{1}{l}{3} & \multicolumn{1}{l}{4} & \multicolumn{1}{l}{5} & \multicolumn{1}{l}{NO QUAL} & \multicolumn{1}{l}{QUAL} & \multicolumn{1}{l}{ALL}  \\ 
\hline
\textbf{\textit{p}} & 0.006                 & 0.233                 & 0.004                 & 0.740                 & 0.033                 & 0.635                 & 0.006                       & 0.003                    & < 0.001                    \\
\textbf{F1}         & 0.609                 & 0.576                 & 0.590                 & 0.594                 & 0.647                 & 0.540                 & 0.591                       & 0.600                    & 0.594                   
\end{tabular}
\end{table}

\subsubsection{Ranking}

One possibility offered by the data captured is the ability to rate samples with respect to the likelihood that a listener might perceive the pieces as human-composed. This could prove a useful feature to researchers whose goal is to understand the qualities of music which people tend to perceive as "human", or to implicitly model them as part of a generative system designed to output increasingly higher-quality symbolic music. Moreover, this can be done in a manner stratified by listener music education level, which may provide further insight into which qualities are preferred or more telling for which people. We use the following equation to obtain such a rating $R$ for any given sample $S_{i} \in \mathcal{G}\cup\mathcal{C}$.

\begin{equation}\label{score}
R_{S_{i}} = N_{S_{i}}^{-0.5} *( \beta_{S_{i}} + \frac{\epsilon}{N_{S_{i}}})
\end{equation}

Equation \ref{score} is designed to balance the absolute ratio of responses deeming the sample to belong to $\mathcal{C}$ with the number of responses received for that particular sample in total, since this ratio becomes more reliably indicative of quality as the number of responses increases. We use $\beta_{S_{i}}$ to denote the number of responses categorising a sample as Bach's music, and $N_{S_{i}}$ for the total number of responses for that sample. We use 0.001 for $\epsilon$, which serves to preserve a logical ranking for any samples where $\beta_{S_{i}} = 0$, by giving those which received more negative responses overall a lower score.

\begin{figure}
\centering
\begin{subfigure}{.5\textwidth}
  \centering
  \includegraphics[width=.7\linewidth]{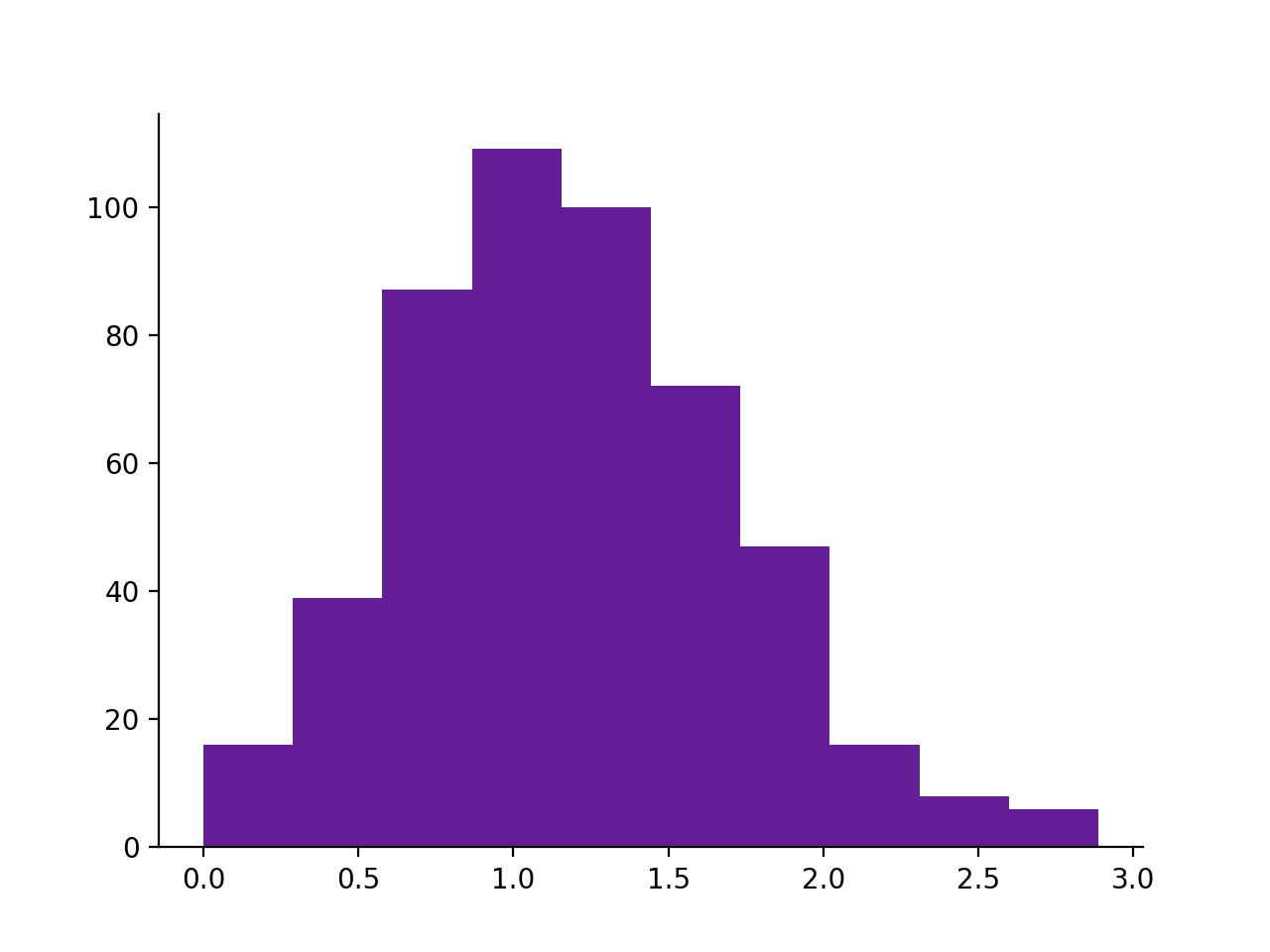}
  \caption{JS Fakes Dataset.}\end{subfigure}%
\begin{subfigure}{.5\textwidth}
  \centering
  \includegraphics[width=.7\linewidth]{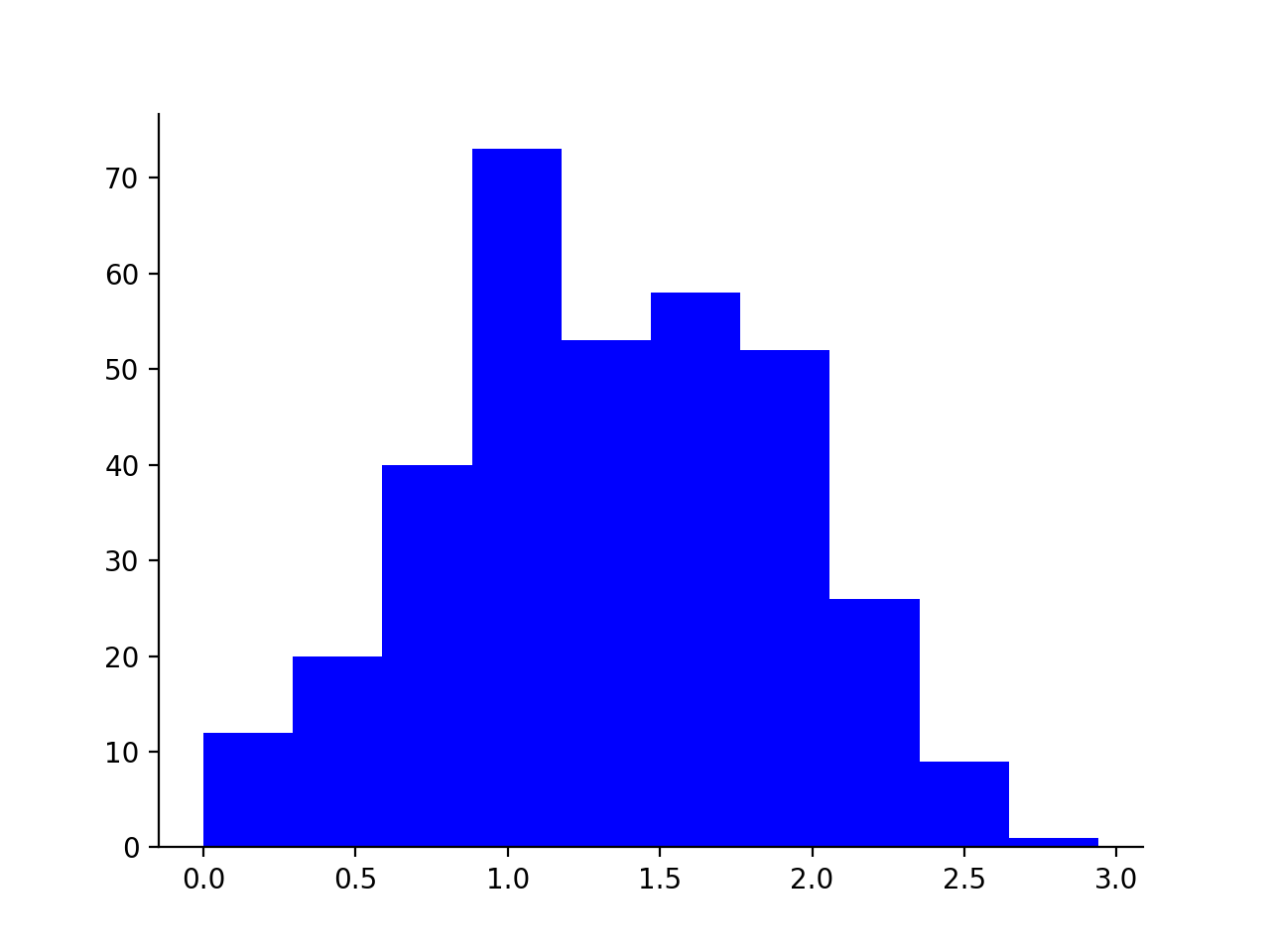}
  \caption{Bach training dataset.}
\end{subfigure}
\caption{Histogram of scores $R$ for pieces in the JS Fakes Dataset (a) and by Bach in the training dataset (b).}
\label{fig:pdf}
\end{figure}

We plot $R$ as histograms for both $\mathcal{G}$ and $\mathcal{C}$ in Figure \ref{fig:pdf}. We can see that samples from $\mathcal{C}$ trend towards slightly higher scores than those from  $\mathcal{G}$, suggesting that pieces by Bach are indeed still more likely to be perceived as human-composed than those by \textit{KS\_Chorus}. Having said that, the distributions of $R$ for the two sets show strong similarities on aggregate. The maximum $R$ for the generated pieces $\mathcal{G}$ is 2.89, while the mean and standard deviation are 1.19 and 0.55 respectively. For $\mathcal{C}$, the maximum, mean and standard deviation are 2.94, 1.33 and 0.57 respectively.

We perform a Mann-Whitney U test on the $R$ scores for the two sets and find $p=0.0493$, which supports our claims that there are similarities between the way the two sets are perceived. Furthermore, we perform the same test on a naive scoring method, for transparency, where the rating is given by the number of "Bach" votes for each sample divided by the total number of votes for that sample. In this case, $p=0.303$. We believe this motivates the use of Equation \ref{score} to calculate rating, as it implicitly factors in a form of confidence by multiplying by a term proportional to $N_{S_{i}}$, offering a more nuanced impression of listener perception.

\subsection{Algorithmic Evaluation}\label{subsec:algoeval}

We conduct experiments to ascertain the benefit of using the JS Fakes Dataset to help with music modelling tasks. We use the canonical JSB Chorales dataset to perform ablation studies, and measure performance in terms of negative log likelihood on the validation set, using the same dataset split as \cite{boulanger}. The task involves element-wise autoregressive modelling of chorales represented in sequence form, as described in Section \ref{subsec:synthesising}. We use the \textit{TonicNet\_Z} algorithm to perform experiments, using the same hyperparameters, training schedule and specific data representation as \cite{tonicnet}. We explore the effect of training only on the JS Fake Chorales and of combining them with the training set of the JSB Chorales as a form of data augmentation. We show how the JS Fake Chorales impact results both with and without augmentation by transposition, and also investigate the impact of limiting the set of JS Fake Chorales seen by the model to those samples with the highest $R$ ratings. Table \ref{tab:tonicnet} shows the results of these experiments, alongside two further strong-performing algorithms.

\begin{table}
\centering
\caption{Results when modelling the J.S. Bach Chorales dataset with two strong baselines and \textit{TonicNet\_Z} trained on various combinations and subsets of the JS Fake Chorales and J.S. Bach Chorales datasets. Rows marked with asterisks are taken directly from \cite{tonicnet}, while rows marked with a cross are taken from \cite{musictransformer}. \newline}
\label{tab:tonicnet}
\begin{tabular}{lllrr}
\textbf{Datasets}       & \textbf{Model}              & \textbf{Aug.}   & \multicolumn{1}{l}{\textbf{Val. NLL}} & \multicolumn{1}{l}{\textbf{NCL Val. NLL}}  \\ 
\hline\hline
Bach\textsuperscript{†} & \textit{CoCoNet (ordered) }           & None            & -                                     & 0.436                                      \\ 
\hline
Bach\textsuperscript{†} & \textit{Music Transformer } & None            & -                                     & 0.335                                      \\ 
\hline
JSF-top                 & \textit{TonicNet\_Z }       & None            & 0.474                                 & 0.363                                      \\
JSF                     & \textit{TonicNet\_Z }       & None            & 0.424                                 & 0.319                                      \\
Bach\textsuperscript{*} & \textit{TonicNet\_Z }       & None            & 0.422                                 & -                                          \\
JSF-top-s               & \textit{TonicNet\_Z }       & Trnsps          & 0.364                                 & 0.272                                      \\
JSF-top                 & \textit{TonicNet\_Z }       & Trnsps          & 0.364                                 & 0.269                                      \\
\textbf{JSF}            & \textit{TonicNet\_Z }       & \textbf{Trnsps} & \textbf{0.328}                        & \textbf{0.234}                             \\
Bach\textsuperscript{*} & \textit{TonicNet\_Z }       & Trnsps          & 0.321                                 & 0.224                                      \\
Bach\textsuperscript{*} & \textit{TonicNet\_Z }       & Trnsps+MM       & 0.317                                 & 0.220                                      \\
Bach+JSF-top            & \textit{TonicNet\_Z }       & Trnsps          & 0.309                                 & 0.215                                      \\
Bach+JSF-top-s          & \textit{TonicNet\_Z }       & Trnsps          & 0.307                                 & 0.213                                      \\
\textbf{Bach+JSF}       & \textit{TonicNet\_Z }       & \textbf{Trnsps} & \textbf{0.300}                        & \textbf{0.208}                            
\end{tabular}
\end{table}

In the first column of Table \ref{tab:tonicnet}, we denote the combinations and subsets of datasets used to train \textit{TonicNet\_Z}. \textit{Bach} here refers to the 229 samples in the JSB Chorales training set and $JSF$ refers to the full set of 500 pieces in the JS Fake Chorales datatet. We also use two special subsets of \textit{JSF} in experiments, namely \textit{JSF-top} and \textit{JSF-top-s}; these represent the subset of \textit{JSF} obtained by ordering the samples by their $R$ scores and taking the top 229. In the case of \textit{JSF-top-s}, we first consider only responses by participants with a skill level of 3 or higher before calculating $R$, to perform an initial investigation into whether insights specifically from individuals with higher levels of musical education might impact downstream results.

The third column shows whether any augmentation techniques were performed on the datasets in question. \textit{Trnsps} refers to transposing the training set only, as far as possible in both direction while all voices in the sample remain within a singable range for their respective voice type. \textit{MM} refers to a crude technique of converting pieces in major keys to minor, and vice versa, as described in \cite{tonicnet}, roughly doubling the dataset size. The remaining columns show experimental results on the validation set, where \textit{NCL} denotes evaluation on the notes only, as is typical in experiments on the JSB Chorales, omitting elements containing chord tokens specifically included as part of the \textit{TonicNet} representation when averaging the loss for each sample.

We can see that even without fine-tuning parameters, training the model only on the JS Fake Chorales obtains results close to training on the original set Bach chorales, whether augmented or not. Remarkably, we are even able to perform better than strong baselines \cite{coconet, musictransformer} without using any real training data. Naturally, there are far more samples in \textit{JSF} than \textit{Bach}, so we use \textit{JSF-top} for a more balanced comparison and find that the model trained on this dataset indeed performs worse, but still within a close range compared to training on the canonical dataset. We see that there is no significant difference between using \textit{JSF-top} and \textit{JSF-top-s} during training, but that using either in combination with \textit{Bach} provides better results than augmenting via \textit{MM}. The dataset size is roughly doubled from the original in all three cases when using \textit{JSF-top}, \textit{JSF-top-s} and \textit{MM}, meaning that the final training set size is close to equal for these cases, in turn suggesting that using the JS Fake Chorales is a higher quality form of augmentation than \textit{MM}. We are therefore able to set a new state-of-the-art result of 0.208 validation set NLL on the JSB Chorales by combining the original training set with the JS Fake Chorales.

\section{Conclusion}

We develop an algorithm capable of generating chorales in the style of Bach and sample 500 pieces from it consecutively with no human intervention. We devise an experiment for human evaluation designed to be fair and free from bias, and through this find that these samples are perceived to be composed by Bach almost as readily as pieces which are in fact by Bach. In so doing, we collect data about human interaction with these pieces and create a dataset of 500 synthetic chorales with human annotation. By avoiding cherry-picking when conducting human evaluation, we demonstrate that the number of pieces in the JS Fake Chorales dataset could be readily scaled while expecting quality to be maintained. We further provide a web interface to readily enable such scaling of the dataset.

We run experiments to demonstrate the effectiveness of this dataset in a common MIR task. We show that training an algorithm on our dataset can be nearly as effective at modelling Bach's music than training on his actual pieces, outperforming several strong baselines. Finally, by combining both the JS Fake Chorales and Bach's chorales into a training corpus, we achieve state-of-the-art results on JSB Chorales dataset.

In general, the motivation for any field of research is to see it lead to real-world applications where it enables people to do what they previously could not, or to to do things with fewer resources than was previously possible. In order to see widespread adoption of generative music algorithms in consumer applications, they must consistently perform well in at least one relevant task. Such tasks in the symbolic music domain include, but are not limited to, melody generation, melody continuation, accompaniment generation, music in-painting and music style transfer. 

Today, no products which enable the aforementioned tasks could truly be described as being widely-used, with the exception of arpeggiators, a family of rule-based tools which perform a specific and narrow kind of style transfer in the symbolic domain. Ultimately this is due to a lack of algorithms which perform with sufficient consistency and quality to provide meaningful value to potential users. One of the most promising ways to overcome this is to break the two bottlenecks mentioned in Section \ref{sec:introduction} and leverage learning algorithms.

Large datasets of music in varying styles, especially containing polyphonic music, will be needed to create learning algorithms that perform all of the above tasks reliably, but these are not easily or cheaply obtainable. Hence we proposed a synthetic dataset with human evaluation data and on-demand scalability as one way to alleviate this problem. Conducting a listening experiment on hundreds of consecutive output allowed us to obtain a good picture of where samples meet or fall short of typical consumer expectation. Newly-generated samples from the same algorithm can be expected to be similarly distributed in this respect, and the function of distance between these samples and consumer expectation, measured as described in Sections \ref{sec:dataset} and \ref{sec:results}, can be modelled in a semi-supervised way in future works. The ability to model this gap gives us a quantitative way to measure it for new algorithms, ultimately allowing researchers to close it more efficiently.

\section{Acknowledgements}

This work was supported by the Polish National Centre for Research and Development under Grant POIR.01.01.01-00-0322/20 titled “Humtap - research and development of the technology for audio-video content generation with machine learning and social sharing”.

\bibliography{jsf_neurips}



\end{document}